\documentclass[aps,prl,twocolumn,showpacs,floatfix]{revtex4}
\usepackage{subfigure}
\usepackage{graphicx}
\usepackage{dcolumn}
\usepackage{epsfig}
\usepackage{amssymb}
\usepackage{amsmath}
\usepackage{rotating}
\usepackage{color}

\begin{document}

\title{Micro-evaporators for kinetic exploration of phase diagrams}

\author{Jacques Leng}\email{jacques.leng@espci.fr}
 \affiliation{ Microfluidique, MEMS and Nanostructures, 
 UMR 7083 CNRS-ESPCI, 10 rue
Vauquelin, 75231 Paris, FRANCE}
\author{Barbara Lonetti}
 \affiliation{ Microfluidique, MEMS and Nanostructures, 
 UMR 7083 CNRS-ESPCI, 10 rue
Vauquelin, 75231 Paris, FRANCE}
\author{Patrick Tabeling}
 \affiliation{ Microfluidique, MEMS and Nanostructures, 
 UMR 7083 CNRS-ESPCI, 10 rue
Vauquelin, 75231 Paris, FRANCE}

\author{Mathieu Joanicot}
 \affiliation{ Rhodia, Laboratory of the Future, 178 av. du Dr. Schweitzer, 33608 Pessac, FRANCE}

\author{Armand Ajdari}
 \affiliation{Physico-Chimie Th\'eorique, UMR 7083 CNRS-ESPCI, 10 rue
Vauquelin, 75231 Paris, FRANCE}

\date{\today}

\begin{abstract}
We use pervaporation-based microfluidic devices to concentrate species in aqueous solutions with spatial and temporal control of the process. Using experiments and modelling, we quantitatively describe the advection-diffusion behavior of the concentration field of various solutions (electrolytes, colloids, etc) and demonstrate the potential
of these devices as universal tools for the kinetic exploration of the phases and textures that form upon concentration.
\end{abstract}
\maketitle

Determination of the phase diagram of multicomponent systems is of importance 
in many realms: industrial formulation, protein cristallization, bottom up material assembly from spontaneous ordering of surfactant, polymeric or colloidal systems~\cite{hbookx,russel,laughlin1999}. 
Depending on the application, one may want to access only the equilibrium phase diagram or gain additional information as to the metastable phases that can appear for kinetic reasons.
Methods to reach these goals often imply tedious and systematic measurements,
requiring for screening purposes the use of robotic platforms. Two generic strategies consist  
in varying 
(in space or time) the temperature of samples of given concentrations on the one hand, and on the other hand isothermal concentration by either removal of the solvent (osmosis, drying), external action on the solutes (sedimentation or dielectrophoresis for colloids), or studies of spontaneous interdiffusion in contact experiments. 

In this Letter we introduce microfluidic tools for controlled isothermal concentration of a wide range of systems, covering solutions of ions, polymers, proteins, surfactants and colloidal suspensions. Our work is inspired by recent observations~\cite{verneuil2004,randall2005} that in standard microsystems built of PolyDiMethylSiloxane (PDMS), {\em spontaneous} water permeation through the PDMS matrix   
 induces flows that can be used to concentrate colloids. Taking a step further, we have engineered specialized microgeometries that allow us to control spatially and temporally the evaporation process as well as the resulting concentration of solutes. Their parallel implementation in microfluidic format
 could open the way to fast screening methods.
 
 After a brief description of the micro-devices, we demonstrate first our control of the concentration process on dilute aqueous solutions of fluorescein and nanoparticles in a simple geometry.
 We then discuss how microfabrication permits to widen the range of possibilities and applications of such devices. As a study case, 
 we report controlled nucleation and growth of crystals of potassium chloride (KCl),
 and show how such experiments provide quantitative information on various thermodynamic quantities (solubility, crystal density) as well as kinetic features (sensitive to the rate of concentration).

\begin{figure}[t]
\begin{center}
\includegraphics[width=7.5cm]{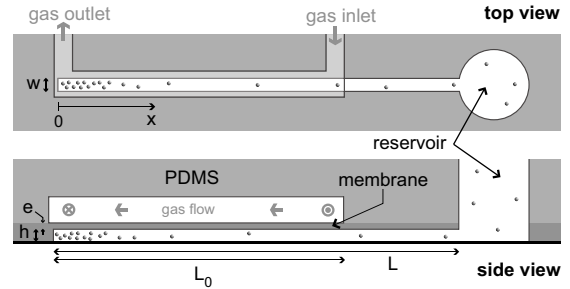}
\end{center}
\vspace*{-7mm}
\caption{\label{fig:setup} Sketch of the finger geometry: top and side views showing the gas and liquid layers and the thin PDMS membrane in between (typical dimensions: $e= 10 ~\mu$m, $h= 20 ~\mu$m,  $w= 200 ~\mu$m, $L_0= 10 ~$mm).}
\end{figure}

{\em The devices -}
The devices used in this paper are two-layer PDMS on glass microsystems (Fig.~\ref{fig:setup}) (fabrication procedure detailed in~\cite{goulpeau2005}).
The microchannels of the bottom layer are filled with the solution of interest,
while air (at controlled humidity) is circulated through the microchannels of the top layer so as to remove the water   
that pervaporates through the thin membrane of PDMS that separates the two networks 
where they overlap (thickness $e$ in the $10-30~\mu m$
range). Many combinations of geometries for the bottom and top networks can be envisaged. We focus here on the simple ``finger'' geometry of Fig. 1, a dead end channel of rectangular cross section (height $h$, width $w$, length $L$) connected to a larger (millimetric) feeding reservoir containing the solution to be concentrated. A terminal section of length $L_0<L$ (typically mms to cms) is covered by the water removal network.

\begin{figure}[t]
\begin{center}
{\includegraphics[width=7.5cm]{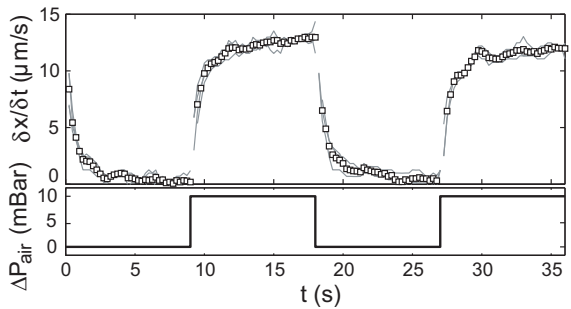}}
\end{center}
\vspace*{-5mm}
\caption{\label{fig:control} Velocity of fluorescent tracers  at a fixed location in a finger in response to the switching on and off of air circulation in the water removal network. Instantaneous velocity (gray lines) is obtained from individual trajectories gathered by Particle Tracking Velocimetry on $1.1~\mu$m diameter tracers. The symbols are for the mean velocity averaged on $\simeq$ 10 trajectories. $L_0 = 12.5$~mm, $h = 22~\mu$m, $e \simeq  20~\mu$m. A latency time of $\approx 0.2$~s is observed before measurements during which the PDMS membrane undergoes vibrations due to the pressure switch.}
\end{figure}

The operation principle is simple: water in the bottom channel 
pervaporates through the thin membrane, which induces a compensating flow
from the reservoir and concentration of the convected solutes at the tip of the finger ~\cite{verneuil2004,randall2005}. 
This is similar to concentration at the boundary of a drying droplet~\cite{deegan1997,clement2004}, without the motion and shear of the concentration zone due to the recess of the liquid-air interface, and without spurious convective flows, thanks to the confinement.
 The small dimensions lead to thermal regulation much faster that typicall kinetics involved in our studies (ms \emph{vs} hours), which permits isothermal studies.

{\em Control of flow and particle concentration -}
To quantify the induced flow,  we adapt the analysis of \cite{verneuil2004,randall2005}, anticipating that evaporation occurs here mostly through the thin membrane, at a volumic flow rate of water $v_e$ (it has dimension of a velocity). Mass conservation then sets the height-averaged velocity in the microchannel: $v(x)=-v_e (x/h)$. Its amplitude rises with the distance $x$ from the dead-end ($x=0$) up to $v_0=v_e(L_0/h)$ at the end of the evaporation zone ($x=L_0$). $v(x)=-v_0$ in the evaporation-free section of the finger $L_0<x<L$. 

Compared to previous work~\cite{verneuil2004,randall2005}, with our dedicated systems we induce larger values of $v_e$ (in the  $50$~nm/s range) due to the thin membranes (permeation yields a limit scaling as $1/e$) and to the dry air flown through the top network (velocities of order cm/s) that reduces the diffusive boundary layer and maintains constant the driving force fpr evaporation. More importantly, we gain a spatial and temporal control on $v_e$ by designing the geometry (evaporation is negligible but for chosen locations) and by tuning in time the air flow and thus $v_e$. 
Quantitative temporal control of the flow field is clear from the motion of tracers at a given location as the air flow is successively turned on and off (Fig. 2).  
When on, tracer velocities of order $v_0 = 13~\mu$m/s are observed, 
corresponding to $\tau_e =h/v_e \simeq  10^3~s$ and $v_e\simeq 22$~nm/s. The velocity drops below $1~\mu$m/s 
when the air flow is turned off, after a response time of a few seconds, compatible with that of the water flux through the thin PDMS layer, $e^2/D_{\rm PDMS}\sim 0.5$~s for $e \simeq 20~\mu$m and a diffusion coefficient for water in PDMS $D_{\rm PDMS}\sim 10^{-9}~$m$^2$/s~\cite{favre1994,watson1996}.

\begin{figure}[t]
\begin{center}
\includegraphics[width=7.5cm]{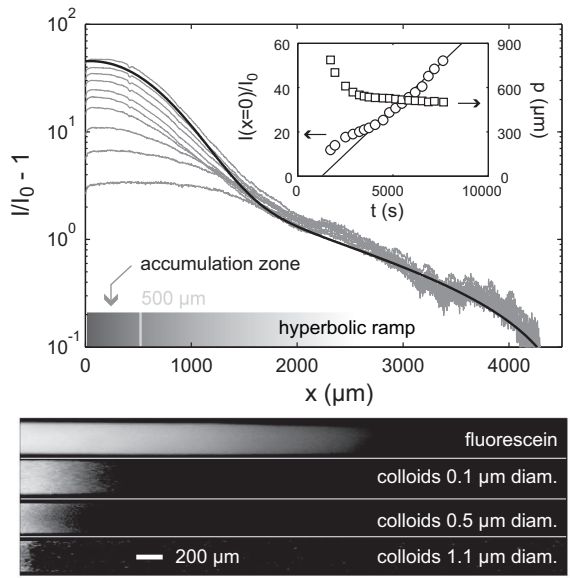}
\end{center}
\vspace*{-3mm}
\caption{\label{fig:fluo} Evaporation-induced concentration. Top: log-linear plot of fluorescence intensity against position at different times
for a finger filled with a aqueous solution of fluorescein. The fit corresponds to the prediction~(\ref{eq:th}). The concentration in the {\em accumulation zone} is well described by a Gaussian hump of width $p \simeq 500~\mu$m, that increases linearly in time (insert), yielding $v_e = 15\pm 2\,$nm/s and $\tau_e = 950 \pm 100$~s. ($c_0= 5\,10^{-5}$~M, $L_0 = 4.7$~mm, $h=15\,\mu$m, $e\simeq 25\,\mu$m; data corrected from photobleaching with a characteristic time of $\approx 2400$~s and total exposure time of $\approx 60$~s). Bottom: images from four fingers in similar conditions showing that the size of the accumulation zone varies with the diffusion coefficient of the markers.}
\end{figure}

Control of the flow field translates into that 
of the induced concentration process. For the simplest case of a dilute species of diffusion
coefficient $D$ in a finger, in a one-dimensional description the conservation equation 
$\partial_t c + \partial_x J=0$ relates the concentration $c(x,t)$ and flux $J(x,t)=cv -D\partial_x c$. 
We focus now on steady evaporation and thus constant $v(x)$ in time,
with a reservoir at fixed concentration $c_0$.
The physics at work is simple: the flow convects the solute towards
the dead end where it accumulates. The current
of solute injected into the finger is steady $J_0= c_0v_0$. Backwards thermal diffusion against the flow controls the width of
the {\em accumulation zone}~\cite{randall2005}, which is $p=(Dh/v_e)^{1/2}=(D\tau_e)^{1/2} <L_0$ 
for strong flows or long fingers $v_0L_o/D \gg 1$ (which we assume from now on~\cite{note1}). At distances larger than $p$, diffusion is negligible and the suspension is simply concentrated by water removal at constant particle flux $c(x)v(x)=c_0v_0$. Altogether, after a transient of duration $\ p^2/D\approx \tau_e$, the profile is well approximated
by a Gaussian (because of the linearity of $v(x)$) increasing linearly in time, ``fed'' by a steady hyperbolic ramp delivering a current $J_0=c_0v_0$:
\begin{equation}
c(x,t) \simeq c_0v_0t  \sqrt{\frac{2}{\pi p^2}} \exp(-\frac{x^2}{2p^2}) + c_0 R(x)
\label{eq:th}
\end{equation}
with $R(x)\simeq L_0/x$ for $p \ll x \ll L_0$. 
Experiments on solutions of fluorescein and nanocolloids quantitatively support this analysis (Fig.~\ref{fig:fluo}).

We thus have multiple ways to reach quantitative control of the kinetics of the process.
Indeed, the concentration at the tip increases as  $\frac{dc}{dt}(x=0)=\sqrt{2/\pi} (v_e/h)^{3/2} c_0L_0 D^{-1/2}$, which shows that control can be exerted 
through geometrical features $L_0$, $h$ and $e$ (that affects $v_e$), 
and through operational parameters $c_0$  and $v_e$.
The latter can be modulated during an experiment, and this allows us 
to pinch or relax concentration profiles at a rate which we have checked to be 
the diffusion-limited response time $\sim p^2/D=h/v_e=\tau_e$ (independent of the species), much larger than the response time of the flow
(typically minutes instead of seconds).

\begin{figure}[t]
\begin{center}
\includegraphics[width=7.5cm]{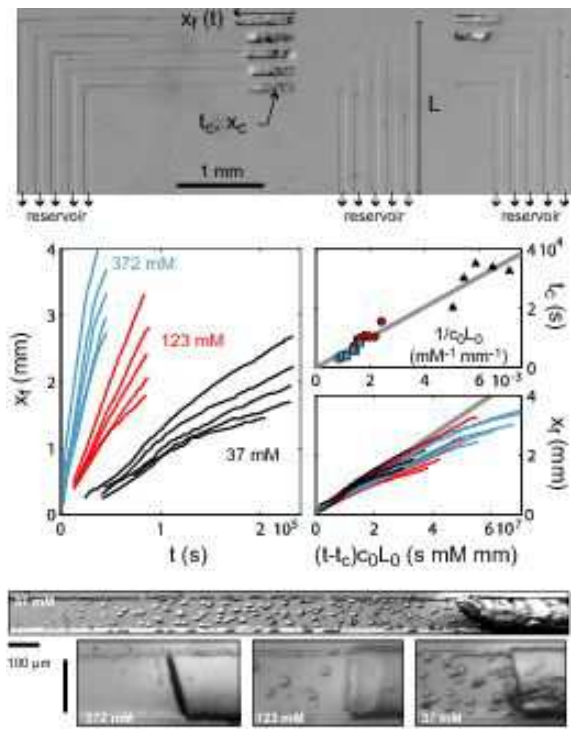}
\end{center}
\vspace{-3mm}
\caption{\label{fig:xtal} 
{\em Top}: view of the chip with 15 microchannels with different evaporation lengths $L_0$ (the bottom of the frame is the end of the evaporation zone). In each channel,  the evaporation-induced concentration leads to crystalization of KCl solutions at $x_c$ after a time $t_c$. Crystal growth is then monitored $x_{f}(t)$.
{\em Middle}:
Left: crystal fronts $x_f(t)$ for various evaporation lengths $L_0$ and initial concentrations $c_0=37$~mM (bottom), 
$c_0=123$~mM (middle), $c_0=372$~mM (top). Right: plots of nucleation time $t_c$ against $1/c_0L_0$, and
front growth $x_f$ against $c_0L_0(t-t_c)$.
{\em Bottom:}
Crystals growing at various $c_0$: ``droplets'' in front of the crystal are only visible at low concentrations.
}
\end{figure}

 {\em Benefits from microfabrication -}
Beyond the simple PDMS finger discussed above, microfabrication offers many  possibilities that we are currently exploring :\\
- We can parallelize tests so as to perform rapid screening with minute amount 
of material. Indeed numerous microchannels of various geometries and types 
can be fabricated on a single chip, and can be fed from a single reservoir or multiple ones. We touch upon the corresponding potentialities in the experiment reported below where a 15-finger chip is used (Fig.~\ref{fig:xtal}). \\
- Geometries are not limited to fingers. Let us illustrate how this can help
on an important example. Equation (1) indicates a drawback of the finger geometry
for the analysis of ill-characterized mixtures
(a drawback actually shared by many other methods : drying, field induced concentration, ...):
different solutes are concentrated at different rates ($p$ depends on $D$) 
so their proportions in the accumulation
zone differ (in an a priori unknown way) from those in the reservoir. 
This can be alleviated by at least two solutions relying on appropriate geometrical design of the channels.
The first is to fabricate very long serpentine fingers and focus
on the hyperbolic concentration ramp ($R(x)$ in equation (1)) 
where all species are concentrated alike.
A second strategy consists in adding a chamber of area $A$ (thickness $h$) at the end of the finger with pervaporation \emph{limited
to the finger}. The area of the accumulation zone is then $\sim A + pw$, and thus essentially independent of $p$ (or $D$) as long as $p\ll A/w$ (with  $A$ a few mm$^2$  and $w$ a few hundred microns, this encompasses anything from ions to micron-sized colloids). Although the concentration rate is lower than in a linear geometry due to the increased accumulation area, the various solutes concentrate at the same rate.  \\
- The PDMS devices presented here will work only with a limited set of solvents that do not swell this elastomer. However microfabrication permits the extension of the concepts presented here 
to other materials. The bottom layer of channels can for example be etched in an impermeable solid matrix (e.g. glass for the ``floor'' and ``side-walls'') separated from the air-circulating network by a thin membrane of a chosen porous materials. Similar sandwich micro-constructions have been used in related contexts ~\cite{russo2004,nama2003,pija2002}.

{\em Controlled crystal nucleation and growth - }
Let us now illustrate how such systems can be used to study physical phenomena that occur upon concentration, by focusing on a well characterized system, KCl aqueous solutions.
We use a microchip with multiple ``fingers'' of different lengths $L_0$ originating from the same reservoir, and perform a few experiments, at the same steady air flux, with different intial concentrations.
Upon concentration, we observe in each finger the nucleation of crystals close to tip, at $x_c$ and $t_c$ (Fig.~\ref{fig:xtal}),
and then their subsequent growth with a front location at $x_f(t)$. The time scales involved vary widely with initial concentration $c_0$ and finger length $L_0$ (Fig.~\ref{fig:xtal}). Nevertheless, a rescaling of
data for both nucleation and growth is possible if we account for the concentration rate proportional to $c_0L_0$ as we now show.

Figure~\ref{fig:xtal} shows that nucleation occurs at a typical time that scales $t_c = K (c_0L_0)^{-1}$, with $K$ a constant. From (1) this corresponds to a nucleation concentration 
$c_c = K \sqrt{2/\pi} D^{-1/2}\tau_e^{-3/2}$. The known values for KCl ($c_c =4.0 \pm 0.5$~M, 
$D \simeq 2.0 \pm 0.2 ~10^{-5}$ ~cm$^2$/s at this temperature) are consistent with the experimental value of $K\simeq 
 5.5 \pm 1 $~sMm and a reasonable value for $\tau_e=h/v_e= 800 \pm 50~$s (which yields $p \approx 1200\,\mu$m).
The rescaling of crystal growth data in Figure ~\ref{fig:xtal} (middle), suggests that the process is limited by the solute feed at $J_0=c_0v_0=c_0L_0/\tau_e$.  Indeed in such a picture, if we assume that the growing crystal at solute concentration $c_X$ fills the channel, then the initial growth rate is dictated by mass conservation
$\frac{dx_f}{dt}= J_0/c_X=  (c_0L_0)/(\tau_e c_X)$. The observed value 
$(c_0L_0)^{-1}\frac{dx_f}{dt} = 0.6\pm 0.1\,10^{-4}$~sMm yields a value $c_X = 20 \pm 2$~M consistent with the literature value (26 M). The observed decrease of the growth rate in time for each channel likely results from that of the evaporation length as the crystal invades the channels.

The above analysis demonstrates a possible use of our devices. Feeding a few fingers with a system of reference (e.g. KCl or fluorescent probes) 
provides a local calibration of the value of $\tau_e$ (to account for variations from device to device). 
Then measurements of $t_c$ and front motion in other fingers fed with the 
solution to analyze yield estimates of $c_c D^{1/2}$ and of the solute concentration $c_X$ in the phase that nucleates.


Further, with our microdevices we can investigate the influence of the kinetics of the concentration process on the morphologies and phases produced. We indeed observe on KCl solutions that both the nucleation and growth processes qualitatively change when the initial concentration $c_0$ (and thus the concentration rate) is varied~\cite{note1}. In the studies reported above,
we reproducibly observe what looks like liquid droplets in the dense electrolyte solution
in front of the growing crystal, but only at the lowest initial concentration examined (Fig.~\ref{fig:xtal} bottom), a phenomenon that we are currently investigating.
Remarkably, the various transient organizations (nucleation scenario, presence or absence of droplets) do not affect the rescalings of Fig. 4, possibly due to the narrow metastable region of KCl and the robustness of mass conservation arguments.  


{\em Perspectives -}
Although we have used KCl to demonstrate some of the potentialities of such microevaporators, 
their main interest likely lies in the rich world of soft matter, where systems adopt many diverse and less known forms of organization~\cite{evans1997,witten1999}.
Beyond ionic solutions and colloids, we are currently performing experiments on surfactant solutions and on proteins~\cite{note1}.
Our early experiments on solutions of Sodium Dodecyl Sulfate (SDS) indeed allow us to observe up to four different mesophases in a given microchannel, and more importantly complex and intriguing dependences of that sequence and of the texture of each phase on the kinetics and history of the concentration process. At the same time, data for nucleation and front growth obey rescalings akin to those of KCl, showing our kinetic control. 

For all the reasons presented above, and especially the delicate control on the kinetics and their wide applicability to  
mixtures of solutes anywhere between simple ions to micron-sized colloids, we believe that our microfluidic evaporators are powerful and rather universal means to explore the wonders of soft matter.


\begin{thebibliography}{16}
\bibitem[{not()}]{note1}
\bibinfo{note}{We postpone to a future work a complete account of our
  advection-diffusion model (with in particular geometries such that the
  accumulation length $p$ is comparable or larger to the length $L_0$ of the
  evaporation zone) and kinetic aspects of concentration-induced phase
  transitions and intermediate states in surfactant and protein solutions.}
\end{thebibliography}


\end{document}